# Resonant Auger Destruction and Iron Kα Spectra in Compact X-ray Sources

Duane A. Liedahl

*Department of Physics and Advanced Technologies, Lawrence Livermore National Laboratory, 7000 East Ave, L-473, Livermore, CA 94550, USA*

**Abstract.** We examine the effects of resonant Auger destruction in modifying the intensities and flux distributions of Kα spectra from iron L-shell ions. Applications include X-ray irradiated stellar winds in X-ray binaries and accretion disk atmospheres. Using detailed atomic models, we find that resonant Auger destruction is selective, in that only a subset of the emitted Kα lines is highly attenuated. We also show that that the local excitation conditions can have a dramatic effect on the Kα emissivity spectrum.

## INTRODUCTION

In high-temperature, highly ionized plasmas, iron K lines, arising typically from *1s-np* transitions, are driven by collisional excitation of H-like and He-like ions. Except under transient ionizing conditions [1] [2], iron K lines from lower charge states contribute relatively subtle modifications to the spectrum [3] [4]. By contrast, in accretion-powered X-ray sources, such as active galactic nuclei (AGN) and X-ray binaries (XRBs), reprocessing of a hard X-ray continuum in relatively cool matter (~$10^5$ – $10^6$ K) can generate intense iron K radiation from more neutral iron species [5].

The reprocessing mechanism, for any species with more than two bound electrons, begins with photoionization of a *1s* electron by a photon of energy $\varepsilon$, sending an element *A* in charge state *i-1* to charge state *i*, where *i* is in a quasi-bound state that is coupled to the continuum, denoted by the double asterisk below.

$$\varepsilon + A_{i-1} \rightarrow A_i^{**} + e^- \quad \text{(K-shell photoionization)}$$

$$A_i^{**} \rightarrow \begin{cases} A_{i+1} + e^- & \text{(autoionization)} \\ A_i^* + \varepsilon_K & \text{(K photon emission)} \end{cases}$$

Autoionization – referred to as the *Auger effect* for *1s* vacancy states – is a possible decay route for the *1s*-hole state. The reaction products are an Auger electron, i.e., an electron with a kinetic energy that is characteristic of the atomic energy level structure, and an ion in charge state *i+1*, where, for the problem at hand, we assume that *i+1* is left in its ground state. Spontaneous radiative decay, the other possible decay route, can stabilize the ion against autoionization. Stabilizing transitions are typically of the parity-changing type *np-1s*, where $n = 2$ for L-shell ions (Kα), and where $n = 2$ or 3 for M-shell ions (Kα or Kβ, respectively). Although the final atomic state is stabilized by a K transition, the final state may be excited, which we denote by the single asterisk.

The two-step process – inner-shell photoionization followed by K line emission – is called *fluorescence*. The *K fluorescence yield,* which we denote by $Y_K$, is the quotient of the rate at which K lines are generated and the rate at which K-shell holes are produced [6]. Once a K-shell hole is produced, $Y_K$ is, for our purposes, determined by the competition between radiative decay and Auger decay. Since electric dipole radiative rates scale as $Z^4$ for $\Delta n > 0$ transitions, and since Auger rates are roughly constant with Z, $Y_K$ increases rapidly with Z [7]. Coupled with the fact that iron has the highest cosmic abundance among intermediate-Z elements, iron K fluorescence lines tend to be the most commonly observed in astrophysics.

Historically, we have had to satisfy ourselves with the fact that spectrometers have, at best, resolved the iron Kα lines into three features: the Fe XXVI Lyα doublet at 6.97 keV, the Fe XXV blend, near 6.7 keV, and a blend of emission from a composite of lines from Fe II-Fe XVIII near 6.4 keV [8]. Readers may note the lack of reference to the ions Fe XIX-XXIV, the subjects of this paper.

For convenience, we refer to the 6.4 keV blend as arising in "near-neutral" material. The blending of lines from near-neutral material [9] [10] can be understood as follows: proceeding through the M shell, successive changes in the screening of the atomic potential experienced by *2p* electrons are small compared to the line energy. Even as we move into the L shell, the successive stripping of M-shell electrons dictates the energy separations. For example, comparing the Kα lines (*1s-2p*) from Fe XVII (Ne-like) and Fe XVIII (F-like), the upper configurations are $1s2s^22p^63s$ and $1s2s^22p^6$, respectively. The small difference in line energy results from the small screening of the nuclear potential by the *3s* electron in the Fe XVII *1s*-hole state. In fact, the Fe XVII and Fe XVIII Kα centroids are separated by only 5 eV. Thus even Kα lines from Fe XVII and Fe XVIII can be thought of as part of the 6.4 keV blend and, in a practical sense, as arising in near-neutral material. By contrast, the Kα energy centroids for Fe XVIII and Fe XIX are separated by 34 eV. This is easily understood by noting that the upper configuration for the Fe XIX lines is $1s2s^22p^5$, where the absence of a *2p* electron has a relatively large effect on the differential screening when compared to the neighboring Fe XVIII. Thus O-like Kα lines are the first to break away from the pack of blended lines from the near-neutral ions. Energy separations of roughly this magnitude persist through the L-shell sequence.

Part of the motivation for this work is the scheduled deployment in 2005 of the XRS calorimeter aboard *Astro-E2*, with a resolution of approximately 6 eV in the iron Kα spectral band [11], which is capable of resolving much of the Kα structure contributed by the L-shell ions. The lack of spectroscopic data with high resolution in this band has undoubtedly played a part in dissuading extensive calculations of iron K spectra. This situation is improving, and detailed calculations of the atomic physics processes associated with iron K spectral formation, using state-of-the-art computer codes, are now becoming available [12] [13].

We have calculated several atomic models in order to search for plasma diagnostics involving iron K spectra. Thus far, the search has focused on the L-shell species, which is a follow-on to similar work on Si Kα spectra (Liedahl, et al., in prep.), which was motivated by an observation of the high-mass X-ray binary Vela X-1 with the gratings aboard *Chandra*. We find that the part of the Si Kα line complex formerly attributed to neutral or near-neutral species [14] is actually made up of features corresponding to Si III – Si XII. The Kα spectra of L-shell ions provide unique diagnostic information on plasmas that exist over intermediate ranges of ionization. It is likely that observations with *Astro-E2* will show that, for some sources, the iron L shell contributes to the overall iron K complex.

Fluorescence from a given ion entails a conversion of part of the radiation field above the corresponding photoelectric edge into discrete emission and Auger electrons. Thus the problem of calculating the fluorescence spectrum from an astrophysical source requires us to consider continuum radiation transfer. Having generated K lines, it makes sense to ask whether or not these lines can escape the fluorescing medium and make their way to an observer. In this paper, we focus on the contributions from discrete opacity to the suppression of photon escape, which involves a process known as resonant Auger destruction, discussed below.

Calculations of atomic structure, radiative transition rates, and autoionization rates and were performed using the HULLAC package [15]. When working with L-shell ions, the stepwise set of processes – photoionization, then autoionization – couples three charge states [16]. In more complex cases involving iron ions of low charge, Auger cascades can couple as many as ten charge states [17]. We will not be concerned with calculations of the charge state distribution here, since we need only consider ``per-atom'' processes.

We assembled six atomic models, one for each of the ions Fe XIX – Fe XXIV. Each model consists of three charge states. For example, in calculating the spectrum of ion *i*, the model also includes charge states *i-1* and *i+1*, where *i-1* serves as a source, through photoionization, of excited states in *i*, and where *i+1* serves as a sink, through autoionization, for these excited states. The treatment of level population kinetics is somewhat involved. Owing to space limitations, a full discussion will be presented elsewhere.

## RESONANT AUGER DESTRUCTION

As with any other photon, discrete opacity can, in principle, suppress emission of fluorescence lines. There is a special circumstance in effect in this case, however: the upper level is autoionizing, which implies a large destruction probability per scatter. The following two-step reaction illustrates the process.

$$\varepsilon_K + A_i^* \rightarrow A_i^{**} \quad \text{(resonant absorption)}$$

$$A_i^{**} \rightarrow A_{i+1} + e^- \quad \text{(Auger decay - photon destruction)}$$

By picking out three of the key words from the above two reactions, one can form the name of the process – *resonant Auger destruction* [18] [19], which we abbreviate RAD. Since the probability of Auger decay is usually high, multiple scattering leading to escape, whether by a random walk in space, or by a migration in frequency to the wings of the line profile, is extremely unlikely.

Clearly, the effect of RAD in attenuating a given line depends on the optical depth in the ion $A_i^*$. Consider two examples in order to make a distinction: Fe XXI (an L-shell ion) and Fe XV (an M-shell ion). For Fe XXI, subsequent to the creation of a $1s$ hole, we have the following set of reactions:

$$1s2s^22p^3 \rightarrow 1s^22s^22p^2 + \varepsilon_K$$
$$\varepsilon_K + 1s^22s^22p^2 \rightarrow 1s2s^22p^3$$

The second reaction is the inverse of the first, except that it involves a distinct ion, since the K photon will travel, on average, a distance corresponding to unity line optical depth from its creation site. The probability of resonant absorption in this case is relatively high, since most of the Fe XXI population will reside in levels belonging to the ground configuration. For Fe XV, we have

$$1s2s^22p^63s^23p \rightarrow 1s^22s^22p^53s^23p + \varepsilon_K$$
$$\varepsilon_K + 1s^22s^22p^53s^23p \rightarrow 1s2s^22p^63s^23p$$

The probability of resonant absorption in this case is relatively low, since the lifetime of $1s^22s^22p^53s^23p$ is exceedingly short, and only a miniscule fraction of the Fe XV population will lie in this configuration. Therefore, as a rule of thumb, RAD does not affect Kα emission from M-shell ions. Note that in the previous example, if Kβ were produced, RAD could efficiently suppress it.

Suppose we adopt a simple picture of a fluorescing zone of an accretion disk in which the zone is represented by a semi-infinite slab that is irradiated from above. Then, based on the preceding discussion, we might expect that only an upper layer of the zone corresponding roughly to unity line optical depth will contribute to Kα fluorescence from L-shell ions. Below that depth, escape is prohibited by resonant Auger destruction. Compare this to the case of fluorescence from M-shell ions, where the fluorescing region for near-neutrals has an extent corresponding to roughly one continuum optical depth. Crudely speaking, therefore, the ratio of the emergent Kα line flux from L-shell ions to the Kα line flux from near-neutrals is of order $\sigma_{cont}/\sigma_{line}$, which is of order $10^{-3}$. This provides a plausible basis for simply setting to zero the Kα line emissivities for L-shell ions, since such low flux levels would imply unobservably small equivalent widths. However, this argument is incomplete, as we discuss below.

# Selectivity

In assessing the effect of RAD, the essential ingredient, apart from the atomic physics, is the line optical depth. For a given transition between an upper level $u$ and a lower level $l$, the line optical depth, in terms of the hydrogen column density $N_H$, the elemental abundance $A_z$, and the ionic fraction $F_{ion}$, can be written

$$\tau_{lu} = N_H A_z F_{ion} \, p_l \, \frac{\pi e^2}{mc} \, f_{lu} \, \phi(\nu)$$

Where $f_{lu}$ is the absorption oscillator strength, $\phi(\nu)$ is the line profile function, and $p_l$ is the fractional population density of the ion. Therefore, we must solve for the level populations $p_l$ for the lower levels of each K$\alpha$ transition. To justify this additional effort, we show in Fig. 1 that K$\alpha$ transitions often terminate on excited states. Under many conditions, for example, in a stellar wind in a HMXB, we expect that the level populations for excited states will usually be much less than for ground, and that plasmas that are optically thick to lines corresponding to ground state transitions need not be optically thick for others. Thus RAD selectively affects the K$\alpha$ spectrum.

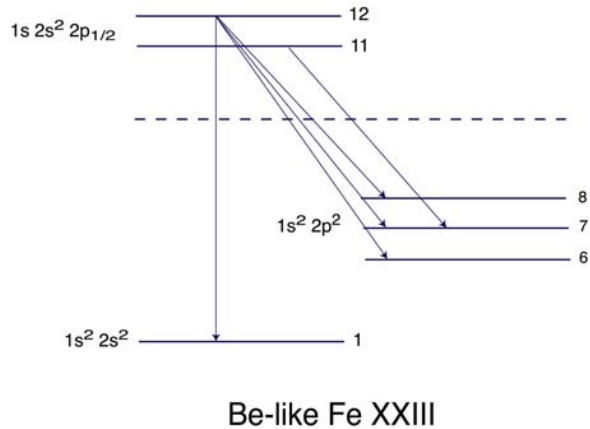

**FIGURE 1.** Partial Grotrian diagram for Be-like Fe XXIII, showing that K$\alpha$ transitions often terminate on excited levels. The numerals indicate the level index for the model ion. The dashed line indicates the first ionization limit.

# Effect of Varying Level Population Distributions

To model iron K$\alpha$ spectra in a disk environment, we add a diluted blackbody radiation field to represent the effect of an underlying accretion disk in driving populations among low-lying excited levels. Once we determine the set $p_l$, we can calculate the line optical depths. Implicit in this discussion is the presence of a hard X-ray field that determines the charge state distribution but does not substantially affect the bound-state level populations. For consistency, we must also calculate the populations for levels in the adjacent ion, the pre-ionization charge state [20]. We note that L-shell ions immersed in plasmas such that the electron densities are near or above the critical densities for $n=2$ intrashell collisional transitions will be similarly affected. For the purpose of illustration, we work with the Be-like/Li-like pair, and determine the Fe XXIV spectrum. We calculate two cases. First, the spectrum is calculated in the limit of low density and no blackbody field. The result is shown in the left panel of Fig. 2. Second, a blackbody with $kT_{rad} = 80$ eV and a dilution factor of 0.5 is added. The result is shown in the right panel of Fig. 2. Here, we are illustrating only the effects of varying the population kinetics – neither example accounts for RAD. The spectrum with the EUV field is characterized by higher energy lines and by brighter Kα emission.

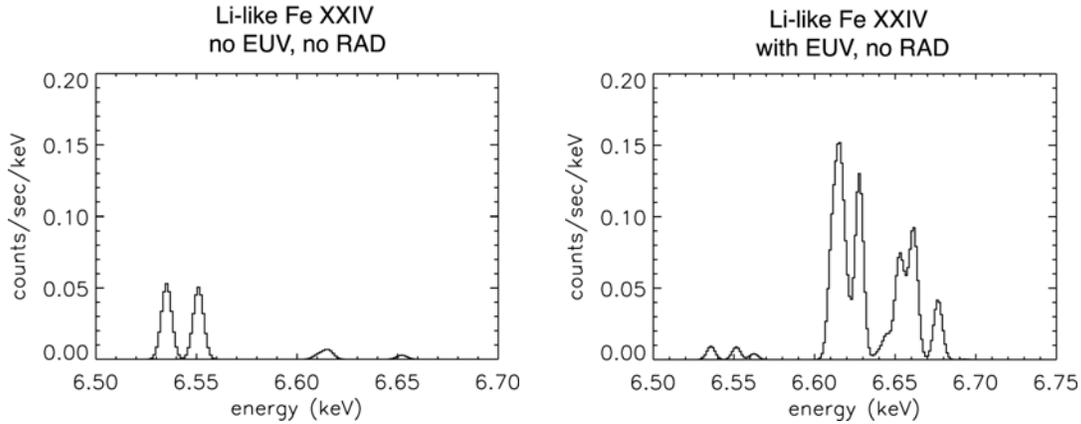

**FIGURE 2.** The left panel shows the Fe XXIV spectrum appropriate to a low-UV environment, in which nearly all charge states exist in their ground levels. The right panel shows the spectrum in the case that the Be-like level population distribution is modified by the presence of a diluted blackbody radiation field. Both examples are "zero-D," i.e., no RAD is applied.

The drastic change induced by application of the Planckian field can be understood by referring to Fig. 3. First, if we assume that nearly all the Be-like population is in the ground state, then the only Li-like autoionizing level produced by K-shell photoionization is $1s2s^2$, which may decay radiatively, as shown by the wavy lines. The line fluorescence yield of each line, i.e., the radiative branching ratio including the total autoionization rate as a sink, is relatively low, as indicated in the figure. When the level populations are re-calculated in the presence of an EUV field, the ground state becomes depleted, while low-lying levels in $1s^22s2p$ gain population. Subsequent K-shell photoionization from this latter configuration populates Li-like $1s2s2p$, which decays by $1s$-$2p$, with a much higher yield.

Next, for a range of ionic column densities appropriate to accretion disk atmospheres [21], we apply RAD to these two emissivity spectra. Using the Voigt function, assuming that the line is the only source of opacity, we calculate an escape probability for each line that is angle-averaged and frequency-averaged. The resulting spectra for the "EUV on" case are shown in Fig. 4, starting with a column density of zero, which represents the spectrum when RAD is neglected. The four panels illustrate the selectivity of the process, and show that, even with moderately large absorbing column densities, RAD does not entirely quench the Kα spectrum.

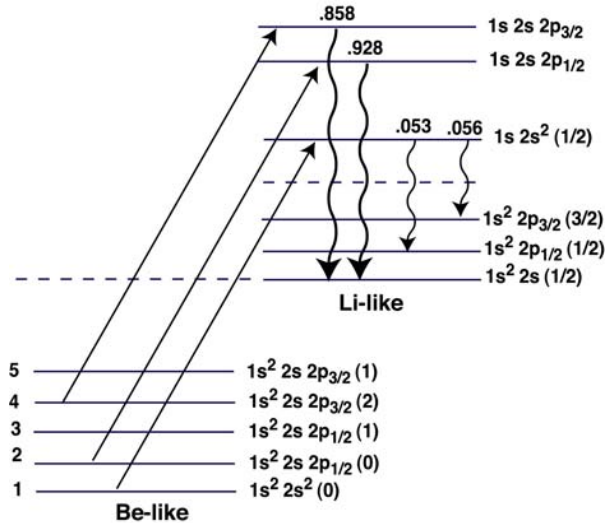

**FIGURE 3.** Schematic of Be-like/Li-like iron showing effect on Fe XXIV Kα spectrum of enhancing populations of low-lying Be-like levels. With the Be-like population concentrated in the ground level, only the *1s2s²* level is driven by K-shell photoionization, which can decay radiatively as shown, though with low line fluorescence yields of about 0.05, shown in the diagram. When Be-like configurations with a *2p* orbital become substantially populated, influx to Li-like *1s2s2p* increases, which has levels with large line fluorescent yields of about 0.9.

A similar calculation was performed for the case of no EUV field. To summarize the results for Fe XXIV, we plot in Fig. 5 (upper left panel) effective fluorescent yields (the apparent fluorescence yield after accounting for RAD) against ionic column density for each of the two cases (EUV on, EUV off). As noted above, the line emissivity for the zero-D case is brighter in the presence of the EUV field than without it. This corresponds to the leftmost side of the plot, where it is seen that the yield is about five times larger than for the "no-UV" case. As the absorbing column density is increased, the effective yield drops, as expected. Note, however, that across this range of column densities, the effective yield is larger than one would calculate assuming that the Be-like population resides exclusively in the ground level. The remaining three panels in Fig. 5 show that Fe XXIV is not unique; the same qualitative behavior is found for Fe XXI, Fe XXII, and Fe XXIII. (Note that, perhaps against convention, that we are assigning the fluorescence yields to the post-ionization charge state, rather than the pre-ionization charge state.) We believe that this casts doubt on the common practice of zeroing out Kα lines from iron L-shell ions.

## CONCLUSIONS

After a re-examination of resonant Auger destruction, where details of the level population distribution are treated explicitly, we find that the effective fluorescent yields for several iron L-shell ions remain high over the range of column densities expected in accretion disk atmospheres. We suggest that models of X-ray irradiated disks can and should be constrained by comparing model predictions of L-shell Kα emission with inferences drawn from observational data. The apparent absence of Kα emission from L-shell ions in accretion disk sources, if verified, for example, by *Astro-E2* observations, may point to problems with the assumptions and/or methodologies adopted for model calculations.

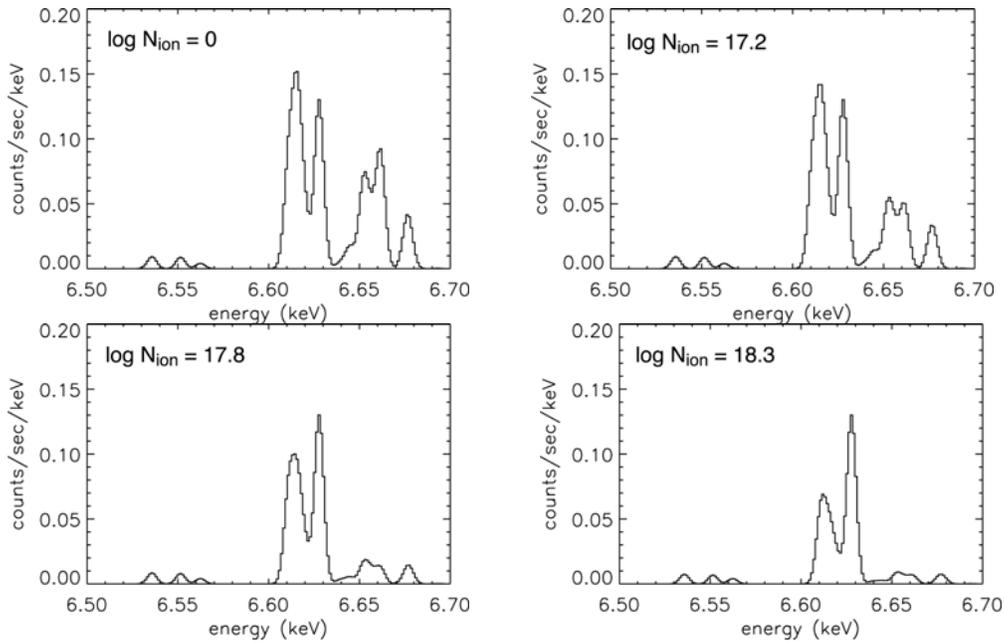

**FIGURE 4.** Fe XXIV spectra plotted for four ionic column densities, showing the modifications engendered by resonant Auger destruction using an escape probability method. The top-left panel shows the Fe XXIV spectrum appropriate to a zero-D calculation. The remaining three show the selective action of RAD. Model spectra are convolved with a 6 eV FWHM gaussian resolution kernel, representative of the *Astro-E2* XRS.

By contrast, we have clear evidence for Kα emission from L-shell ions in HMXBs. However, these detections are not of iron emission, but rather, silicon and sulfur emission, and we are not aware of any reports of this emission from L-shell iron. While this apparent discrepancy has not been studied in detail, it does appear to pose an interesting puzzle. The problem will be better defined by results from *Astro-E2*. In HMXBs, the RAD effect may turn out to be a useful diagnostic based on analyses of line ratios or "line" shapes, recalling that most ionic "lines" are blends. Clearly, having observational access to a class of lines from several elements that may respond differently to RAD, and that respond differently to local excitation conditions, offers an exciting opportunity to gain insight as to the composition, physical state, and dynamics of accretion flows in these systems. On the other hand, the application of specific diagnostics based on RAD in relativistic accretion disks, where the detail is blurred into a single, though possibly structured, line complex, is not likely. There, it is the energy distribution and total equivalent width that matter most. Nevertheless, these gross attributes depend on detailed modeling calculations.

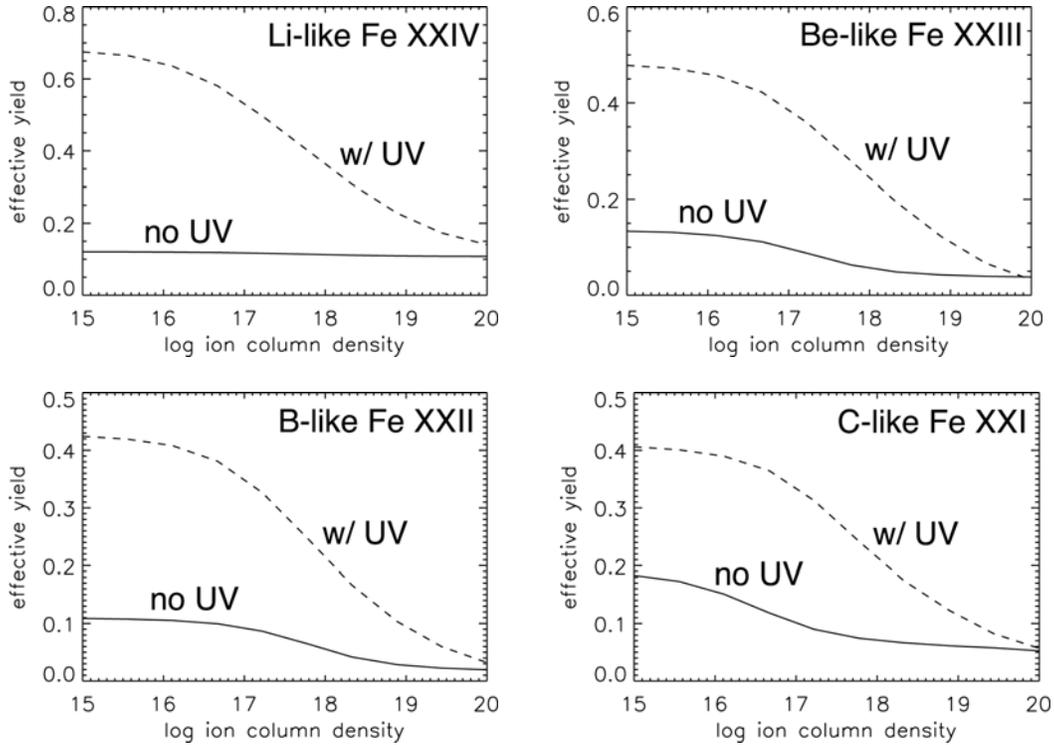

**FIGURE 5.** Effective fluorescent yields plotted against ionic column density for Fe XXI – Fe XXIV. Each plot shows two cases: (dotted line) level populations of low-lying states driven by 80 eV half-diluted blackbody and (dashed line) no perturbing radiation field.

We acknowledge the inadequacy of the escape probability treatment in assessing spectral modifications to the zero-D emissivity spectrum. The underlying assumption in using an escape probability formalism is that a photon, once resonantly absorbed, is destroyed via the Auger effect. This is not a horrible approximation, since the destruction probabilities are, on average, rather high. But the distribution in line fluorescent yields is broad (see Fig. 3, for example), and multiple scattering leading to escape cannot be dismissed outright. Moreover, we have ignored line splitting (see Fig. 1), where the photon energy is converted to one corresponding to a different energy level separation, such that the lower level is less populated than the lower level corresponding to the pre-absorption line, thereby increasing the probability of escape. Finally, we assumed a homogeneous slab in calculating spectral modifications for different column densities, whereas a more reliable calculation would account for density, temperature, and ionization gradients, and ultimately, the effects of velocity fields. These aspects of the problem will be taken up in a future paper, using the Monte Carlo radiation transport capability of the COMPASS code [22].

## ACKNOWLEDGMENTS

The author thanks Mau Chen, Tim Kallman, and Chris Mauche for useful discussions. This work was performed under the auspices of the U.S. Department of Energy by the University of California Lawrence Livermore National Laboratory under contract No. W-7405-Eng-48.